\documentclass{elsart}
\usepackage{amssymb}
\usepackage{amsmath}
\usepackage[dvips]{graphicx}

\journal{:\bf \   Comptes Rendus Acad. Sci. Paris   II b, t. 320,
n$^o$ 5, p. 211-216\
(1995).\qquad\qquad\qquad\qquad\qquad\qquad\qquad\qquad\qquad\qquad\qquad
}
\input{tcilatex}
\begin{document}

\begin{frontmatter}

\title{Radius and surface tension of microscopic bubbles by second gradient
theory}

\author{Francesco dell'Isola}
\address {
Universit\`a degli Studi di Roma La Sapienza, Dipartimento di
Ingegneria  Strutturale e Geotecnica, Via Eudossiana, 18 - 00184
Roma Italy} \ead{isola@scilla.ing.uniroma1.it}
\author{Henri Gouin}

\address {
C.N.R.S.  U.M.R. 6181 \&  Universit\'e d'Aix-Marseille \\ Case 322,
Av. Escadrille
 Normandie-Niemen, 13397 Marseille Cedex 20 France}
 \ead{ henri.gouin@univ-cezanne.fr}
\author{Pierre Seppecher}
\address {
Institut de Math\'ematiques de Toulon, Universit\'e de Toulon et du
Var
\\ Avenue de l'Universit\'e, BP 132, 83957 La Garde Cedex  France}
\ead{seppecher@imath.fr}
\begin{abstract}
The classical theory of Laplace is not suitable for describing the
behavior of microscopic bubbles. The theory of second gradient
fluids (which are able to exert shear stresses in equilibrium
conditions) allows us to obtain a new expression for surface tension
and radius of these bubbles in terms of functionals of the chemical
potential. This relationship allows us to generalize the results of
Cahn-Hilliard (1959) and Tolman (1948).
\end{abstract}


\end{frontmatter}

\section{Introduction}

The microscopic bubbles we consider in this paper contain enough
molecules to be modelled as a continuous strongly non-homogeneous
system. To the classical expression for the free energy valid for
homogeneous continua has to be added a term depending on the
variation of mass density.\ Usually the additional term is assumed
to depend on the gradient of density only (Cahn, 1959; Serrin,
1986). Hence the expression for stress tensor differs from that
valid for elastic fluids as even in equilibrium conditions shear
components also appear. The form of the balance equation of force
valid for second gradient fluids is found in Germain (1973), Casal
(1962). In 1948, Tolman established, by using a Gibbs-like approach
to Laplace theory, some relationships between the radius and the
surface tension of bubbles in equilibrium with their liquid phase.\
This work produced many theoritical and experimental investigations
concerning bubbles of quasi-molecular dimensions (Fisher, 1980;
Kumar,1991). In this case the theory of Laplace is not suitable as
the vapour bubble consists mainly or exclusively of the interfacial
phase. However, to interpret the experimental evidence it is still
important to attribute a radius and an energy to microscopic germs:
we do this by means of an \textit{equivalent }model of Laplace type.
We note that Cahn and Hilliard did not consider the mechanical
aspect of the
nucleation phenomenon. They only considered the thermodynamic pressure (%
\textit{i.e.} the spherical part of stress tensor deriving from the
classical expression for free energy) while in Casal (1985) is
shown, for second gradient fluids, the existence of a capillary
non-spherical stress tensor whose trace includes - but does not
reduce to - quoted thermodynamic pressure. The analysis of some
preliminary numerical calculations allows us to conclude that the
theory of second gradient fluids, reinterpreted through a comparison
with Laplace theory, leads to predictions closer to the experimental
evidence than those available in the literature for radii close to
the critical one. A further improvement is conceivable by
considering non-constant capillarity coefficient as suggested by De
Gennes (1981).
\section{Equilibrium of bubbles and Gibbs rule by the second gradient theory}
The balance equations of force for capillary fluids are available in the
literature (\textit{see }e.g. Serrin, 1986).\ The most simple model taking
into account the non-homogeneity effects introduces a unique additional
physical constant $C$ (Casal \& Gouin, 1985). Its value in mks units is very
small: its effects are notable only inside interfaces. Balance of forces, in
equilibrium conditions, reads ($\tau ,\rho ,\Omega $ denoting stress tensor
of non-dissipative capillary fluids, mass density and potential of body
forces):
\begin{equation}
\func{div}\tau -\rho \func{grad}\Omega =0,
\end{equation}%
with%
\begin{equation*}
\begin{array}{lll}
\tau =-p~\text{Id}+C\func{grad}\rho \otimes \func{grad}\rho , & \text{where}
& \displaystyle p=\mathcal{P}-C\frac{(\func{grad}\rho )%
{{}^2}%
}{2}+C\rho ~\Delta \rho ,%
\end{array}%
\end{equation*}%
with denoting by $\mathcal{P}$ and $\Delta $ the thermodynamic pressure and
the Laplace operator respectively. Equation (1) implies:%
\begin{equation}
\func{grad}\mathcal{P}+\rho \func{grad}(\Omega -C~\Delta \rho )=0.
\end{equation}%
With assuming isothermal equilibrium conditions,  relation (2)
becomes:%
\begin{equation}
\mu (\rho )-C~\Delta \rho +\Omega =C^{ste},
\end{equation}%
where $\mu $ is the specific free enthalpy (\textit{i.e.} the chemical Gibbs
potential) relative to the fluid in a homogeneous state with mass density $%
\rho $ and temperature $T$. If $\Omega $ is negligible, the equilibrium of a
bubble surrounded by its liquid phase with density $\rho _{l}$ is
represented by a spherically symmetric profile of density satisfying
(Rocard, 1952; Cahn \& Hilliard,1959):%
\begin{equation}
C\frac{d%
{{}^2}%
\rho }{dr%
{{}^2}%
}+2\frac{C}{r}\frac{d\rho }{dr}=\mu (\rho )-\mu (\rho _{l}).
\end{equation}%
Equation(4) has to be supplemented with appropriate boundary conditions:

- \textit{(a) }because of spherical symmetry, the derivative of\textit{\ }$%
\rho $ vanishes at the origin,

- \textit{(b) }as we have assumed that the bubble is surrounded by a
homogeneous liquid, the derivative of\textit{\ }$%
\rho $ vanishes  at infinity.

If $\rho _{l}^{P}$ is the density of the liquid phase in the
equilibrium state with plane interface and the function $\mu $ is
smooth enough, then the theory of Fuchs equations (Valiron) implies
that:

(i) for every $\rho _{l}$ in an interval $\left] \rho _{m},\rho _{l}^{P}%
\right[ $, Eq. (4) and conditions \textit{(a)-(b)} uniquely
determine an increasing mass density profile $\rho (r)$ and in
particular its value $\rho _{v}$ at the origin: we say that $\rho
(r)$ satisfies the capillary fluids version of the \textit{Gibbs
Phase Rule};

(ii) the function $\rho $ is twice differentiable at the origin;

(iii) as $\displaystyle \frac{d\mu }{d\rho }\rho _{l}> 0$, $\rho $
converges at least exponentially to $\rho _{l}$ when $r$ tends to
infinity.\ With denoting with a prime the radial derivative and by
$\psi $ the free energy per volume, Eq. (4) implies:

\begin{equation}
C\rho ^{\prime \prime }\rho ^{\prime }+\frac{2C}{r}\rho ^{\prime }%
{{}^2}%
=\mu (\rho )\rho ^{\prime }-\mu (\rho _{l})\rho ^{\prime },
\end{equation}%
and by integrating
\begin{equation}
\frac{C}{2}\rho ^{\prime }%
{{}^2}%
+\int_{0}^{r}\frac{2C}{x}\rho ^{\prime }%
{{}^2}%
(x)~dx=\psi (\rho )-\psi (\rho _{v})+\mu (\rho _{l})(\rho _{v}-\rho
).\end{equation}

\section{Nucleation energy of bubbles}

We deal here with bubbles which are small with respect to a typical
size of the liquid phase. More precisely, a mass density field $\rho
,$ whose mean value is $\rho _{0},$ in a domain $\mathcal{D}$ of
volume $v(\mathcal{D)}$ represents a small bubble if $\rho _{l}$ and
$\varepsilon \ll 1$ exist such that $B=\{x\in \mathcal{D\diagup
}\displaystyle \left\vert \frac{\rho (x)-\rho _{l}}{\rho
_{0}}\right\vert
> \varepsilon \}$ satisfies $\displaystyle\frac{v(B)}{v(\mathcal{D)}}<
\varepsilon$. We note that, as $\displaystyle
\int_{\mathcal{D}}(\rho -\rho _{0})~dV=0,$ $\displaystyle\left\vert \frac{%
\rho _{0}-\rho _{l}}{\rho _{0}}\right\vert =O(\varepsilon )$. We now
evaluate (in the absence of capillarity effects and at the first
order of approximation in $\varepsilon $) the difference between the
free energy of such a small bubble configuration and the homogeneous
configuration of density $\rho _{0}$.\ This difference, which is
part of the nucleation energy of the bubble, is $\displaystyle
w=\int_{\mathcal{D}}(\psi (\rho )-\psi (\rho _{0}))~dV.$

As $\displaystyle \int_{\mathcal{D}}\rho _{0}~dV=\int_{\mathcal{D}}\rho ~dV,$
the energy $w=\displaystyle\int_{\mathcal{D}}[\psi (\rho )-a\rho -\psi (\rho
_{0})+a\rho _{0}]~dV$ does not depend on the choice of the constant $a$. The
total energy is well defined but its localization is somewhat arbitrary.\
This indetermination is related to the arbitrariness in the choice of
chemical potentials.\ With the choice $a=\mu (\rho _{l})$ the energy is
localized inside $B$.\ Indeed, with using the previous estimation for $\rho
_{0}-\rho _{l}$ we obtain:%
\begin{equation*}
w=\int_{B}[\psi (\rho )-\mu (\rho _{l})(\rho -\rho _{l})-\psi (\rho
_{l})]~dV+O(\varepsilon
{{}^2}%
).
\end{equation*}%
To the non-capillary part of nucleation energy of a bubble in an unbounded
domain we have to add the interfacial (capillary) energy. In the
Gibbs-Laplace theory we have for bubbles of radius $R$ and surface tension $%
\sigma $:%
\begin{equation}
W=4\pi R%
{{}^2}%
\sigma +\frac{4}{3}\pi R^{3}[\psi (\rho _{v})-\psi (\rho _{l})+\mu (\rho
_{l})(\rho _{l}-\rho _{v})].
\end{equation}%
In the theory of second gradient fluids, following Cahn and Hilliard, we
have:%
\begin{equation}
W=\int_{\mathcal{D}}[\psi (\rho )-\psi (\rho _{l})-\mu (\rho _{l})(\rho
-\rho _{l})+\frac{C}{2}(\func{grad}\rho )%
{{}^2}%
]~dV.
\end{equation}%
Let $\mathcal{P=}\rho \mu -\psi $ denote the thermodynamic pressure. The
conditions%
\begin{equation}
\mu (\rho _{l})=\mu (\rho _{v}),
\end{equation}%
\begin{equation}
\mathcal{P}(\rho _{l})-\mathcal{P}(\rho _{v})\mathcal{=-}\frac{2\sigma }{R},
\end{equation}%
valid - for isothermal equilibrium - only in Laplace theory, transform
Eq.(7) into:%
\begin{equation}
W=\frac{4}{3}\,\pi R%
{{}^2}%
\sigma,
\end{equation}%
and imply that the nucleation energy of the bubble is the third of the
creation energy of its interface.\ We can extend this result to the theory
of second gradient fluids.\ Let $\phi (\rho )=\psi (\rho )-\psi (\rho
_{l})-\mu (\rho _{l})(\rho -\rho _{l});$ by multiplying Eq. (6) by $r%
{{}^2}%
$ and integrating it over $[0$,$\infty ]$ we get:
\begin{equation}
\int_{0}^{\infty }r%
{{}^2}%
\left( \frac{C}{2}\rho ^{\prime }%
{{}^2}%
-\phi (\rho )\right) ~dr+\int_{0}^{\infty }r%
{{}^2}%
\left[ \phi (\rho _{v})+\int_{0}^{r}\frac{2C}{x}\rho ^{\prime }%
{{}^2}%
(x)~dx\right] ~dr=0,
\end{equation}
by integrating by parts and using Eq. (6) again, we obtain:%
\begin{equation}
\int_{0}^{\infty }r%
{{}^2}%
\left( \frac{C}{6}\rho ^{\prime }%
{{}^2}%
+\phi (\rho )\right) ~dr+\left[ \frac{r^{3}}{3}\left( \phi (\rho )-\frac{C}{2%
}\rho ^{\prime }%
{{}^2}%
\right) \right] _{0}^{\infty }=0.
\end{equation}%
Because of (i) - (iii) of Sect. 2, the last term vanishes so that we obtain%
\begin{equation}
W=4\pi \int_{0}^{\infty }r%
{{}^2}%
\left( \frac{C}{2}\rho ^{\prime }%
{{}^2}%
+\phi (\rho )\right) ~dr=\frac{4}{3}\pi \int_{0}^{\infty }Cr%
{{}^2}%
\rho ^{\prime }%
{{}^2}%
~dr.
\end{equation}%
With denoting by $R_{m}^{%
{{}^2}%
}$ the mean value of $r%
{{}^2}%
$ with respect to the measure $\rho ^{\prime }%
{{}^2}%
dr$, Eq.(14) reads%
\begin{equation}
W=\frac{4}{3}\,\pi R_{m}^{2}\int_{0}^{\infty }C\rho ^{\prime }%
{{}^2}%
~dr.
\end{equation}%
For a large enough bubble (\textit{i.e. }when\textit{\ }$\rho _{l}$ tends to
$\rho _{l}^{P})$:

(i) $R_{m}$ is the radius,

(ii) Equation (15) reduces to Equation (11), as surface tension for plane
interface is $\displaystyle
\int_{0}^{\infty }C\rho ^{\prime }%
{{}^2}%
~dr.$

\section{Comparison between Laplace and second gradient theories. Equivalent
bubbles}

In second gradient theory the stress tensor in the center of a spherical
bubble takes the value $\tau =-p_{v}Id$\, where $p_{v}=\mathcal{P}(\rho _{v})
$. As Equation (4) implies $C\rho _{v}\Delta \rho _{v}=\rho _{v}\left( \mu
(\rho _{v})-\mu (\rho _{l})\right) $, we have%
\begin{equation}
p_{v}-p_{l}=\psi (\rho _{l})-\psi (\rho _{v})+\mu (\rho _{l})(\rho _{v}-\rho
_{l}).
\end{equation}%
Let us note that this difference is not equal to the corresponding
difference of thermodynamic pressures as, for microscopic bubbles, $\mu
(\rho _{l})$ differs from $\mu (\rho _{v}).$ As experimental results
(Fisher, 1980) deal with measures of stresses, then we have to use $%
p_{v}-p_{l}$ instead of $\mathcal{P}(\rho _{v})-\mathcal{P}(\rho _{l})$ in
the comparison between Laplace and second gradient theories. We can now
define the surface tension and the radius of a bubble by identifying the
nucleation energies and the pressure differences computed in both theories.\
Indeed $\displaystyle
p_{v}-p_{l}=\frac{2\sigma }{R}$ and $\displaystyle \frac{4}{3}\pi
\int_{0}^{\infty }Cr%
{{}^2}%
\rho ^{\prime }%
{{}^2}%
~dr=\frac{4}{3}\pi R%
{{}^2}%
\sigma $ imply%
\begin{equation}
R=\left[ 2C\int_{0}^{\infty }r%
{{}^2}%
\rho ^{\prime }%
{{}^2}%
~dr\right] ^{\frac{1}{3}}\left[ \psi (\rho _{l})-\psi (\rho _{v})+\mu (\rho
_{l})(\rho _{v}-\rho _{l})\right] ^{-\frac{1}{3}},
\end{equation}
and
\begin{equation}
\sigma =\left[ \frac{C}{4}\int_{0}^{\infty }r%
{{}^2}%
\rho ^{\prime }%
{{}^2}%
~dr\right] ^{\frac{1}{3}}\left[ \psi (\rho _{l})-\psi (\rho _{v})+\mu (\rho
_{l})(\rho _{v}-\rho _{l})\right] ^{\frac{2}{3}}.
\end{equation}

\section{Conclusion}

Let us notice that our treatment is based on Eq. (1) obtained by a
continuum model.\ However, van Kampen (1964) obtained the same
differential equation by using ideas and methods of statistical
mechanics.\ Eq. (14) has allowed us to evaluate the nucleation free
energy of a bubble, once the mass density profile is known,
\textit{i.e. }once the liquid phase density $\rho _{l}$ has been
assigned. Equation (16) allows us to understand the difference
between thermodynamic and stress pressure.\ These two equations do
not depend on a particular constitutive law and determine the radius
and surface tension for a microscopic bubble.\ This is done by
identifying the pressure jump and nucleation energy of Laplace
theory with the corresponding quantities in the second gradient
theory.\ This method has never been used in literature: for example,
Cahn (1959) and Evans (1979) used for surface tension the expression
valid for plane interfaces, and did not investigate the relationship
between surface tension and radius.\ Our Eqs. (17)-(18) give such a
relationship in an implicit form: in fact, both surface tension and
radius depend on the density profile, which, in turn, depends on
$\rho _{l}$. When the radius tends to infinity, our expression for
surface tension reduces to that for plane interfaces.\ We observe
that the relationship between $\sigma $ and $R$ depends on
thermodynamic potential. Numerical calculations were performed by
Dell'Isola, Gouin and Rotoli (1996) using   equations (17)-(18) and
van der Waals-type potential proposed by van Kampen (1964), Rocard
(1967) and Peng (1976). It was able to observe that, for bubbles
whose radius is close to the critical one, the predicted variations
of the surface
tension (for water and cyclohexane at 20$%
{{}^\circ}%
C)$ are in good agreement with the experimental observations (Katz \textit{%
et al }1976; Fisher \& Israelachvili, 1980).\ This is not the case for the
theoretical treatment proposed by Tolman (1947) and Kumar \textit{et al}
(1991).

{\bf References}

{ \footnotesize
 Cahn J.W., Hilliard J.E., Free energy of a non uniform system III,
J. Chem. Phys. \textbf{31}, pp. 688-699 (1959).

 Casal P., La capillarit\'{e} interne, Cahier du groupe Fran\c cais de
rh\'{e}ologie, CNRS VI, \textbf{3}, pp. 31-37 (1961).

 Casal P., Gouin H., Connection between the energy equation \ and the
motion equation in Korteweg's theory of capillarity, C. R. Acad.
Sci. Paris, \textbf{300}, S\'{e}rie II, pp. 231-234 (1985).

 de Gennes P.G., Some effects of long range forces on interfacial
phenomena, J. Physique-Lettres \textbf{42}, L-377, L-379 (1981).

Dell'Isola  F., Gouin H., Rotoli G., Nucleation of spherical
shell-like interfaces by second gradient theory: numerical
simulations, Eur. J. Mech., B/Fluids, \textbf{15}, 4, pp. 545-568
(1996).

  Evans R., The nature of the liquid-vapor interface and other
topics in the statistical mechanics of non-uniform, classical fluids, Adv.
in Phys., \textbf{28}, pp. 143-200 (1979).

  Fisher L.R., Israelashvili J.N., Determination of the capillary
pressure in menisci of molecular dimensions, Chem. Phys. Letters,
\textbf{76}, pp. 325-328 (1980).

  Germain P., La m\'{e}thode des puissances virtuelles en m\'{e}%
canique des milieux continus, J. de M\'{e}canique, \textbf{12}, pp. 235-274
(1973).

  Gibbs J.W., Collected works, 1, Yale Univ. Press, 1948.

  Katz J.L., Mirabel P., Scoppa C.J. and Virkler T.L.,
Condensation of a supersatured vapor III.\ The homogeneous nucleation of CCl$%
_{4},$CHCl$_{3}$, CCl$_{3}$F and C$_{2}$H$_{2}$Cl$_{4}$, J. Chem.
Phys.,  \textbf{65}, pp. 382-392 (1976).

  Kumar F.J., Jayaraman D., Subramanian C., Ramasamy P.,
Curvature dependence of surface free energy and nucleation kinetics of CCl$%
_{4}$ and C$_{2}$H$_{2}$Cl$_{4}$ vapours, J. of Material Sci. Lett., \textbf{%
10}, pp. 608-610 (1991).

  Peng D., Robinson D.B., A new two-constant equation of state,
Ind. Eng. Chem. Fundam., \textbf{15}, pp. 59-64 (1976).

  Rocard Y., {Thermodynamique}, Masson, Paris, Chapter \textbf{\textbf{V}},
1967.

  Serrin J., {New perspectives in thermodynamics},
Springer Verlag, Berlin, New York, pp. 187-260 (1986).

  Tolman R.C., Consideration of the Gibbs theory of surface
tension, J. Chem. Phys., \textbf{16}, pp. 758-774 (1948).

  Valiron G., Equations fonctionnelles, applications, Masson,
Paris, 1950.

  Van Kampen N.G., Condensation of a classical gas with long
range attraction, Phys. Rev., \textbf{135}, A362-A369 (1964).}

\vskip 1cm \centerline{\bf
---------------------------------------------------------------------------------------}
\vskip 1cm

\centerline {\bf Abridged French version} \vskip 0.2cm \centerline
{\bf Rayon et tension superficielle des bulles microscopiques}
 \centerline
{\bf  en th\'eorie du second gradient} \vskip 0.2cm {\small
\centerline { {\textbf{R\'esum\'e} } }

{\small La th\'eorie de Laplace est inadapt\'ee pour \'etudier les
bulles de dimensions mol\'eculaires. La th\'eorie des fluides
dou\'es de capillarit\'e interne nous permet de proposer une
expression de la tension superficielle et du rayon des bulles comme
fonctionelles du potentiel chimique. Cette expression, en accord
avec l'exp\'erience, am\'eliore les r\'{e}sultas obtenus par
Cahn-Hilliard et Tolman. \vskip 0.5cm }

{\small \centerline{\bf Version fran\c caise  abr\'eg\'ee} }

{\small Pour \'{e}tudier le comportement d'un fluide fortement h\'{e}t\'{e}%
rog\`{e}ne on ajoute \`{a} l'expres-sion de la densit\'{e} d'\'{e}nergie
libre d'un fluide homog\`{e}ne un terme limit\'{e}, le plus souvent, \`{a}
un d\'{e}veloppement au second ordre en gradients (Cahn, 1959; Serrin,
1986). Les \'{e}quations du mouvement de ces fluides sont obtenues par la th%
\'{e}orie du second gradient (Casal, 1961, Germain, 1973).\newline
Tolman (1948) a \'{e}tudi\'{e} la tension superficielle des bulles
en fonction de leur rayon. Des mesures exp\'{e}rimentales ont
\'{e}t\'{e} faites pour des bulles de dimensions quasi
mol\'{e}-culaires (Fisher, 1950; Kumar 1991). Pour de telles
dimensions la th\'{e}orie de Laplace est disqualifi\'{e}e car la
bulle de vapeur devient l'interface elle-m\^{e}me. Il est
n\'{e}anmoins important de donner une dimension et une \'{e}nergie
pour les germes microscopiques puis de les comparer \`{a} un
mod\`{e}le \'{e}quivalent de
type Laplace. On peut ainsi interpr\'{e}ter les r\'{e}sultats des exp\'{e}%
riences.\ Le mod\`{e}le le plus simple de fluide du second gradient ne fait
intervenir qu'une seule constante physique suppl\'{e}mentaire\ }${\small C}$%
{\small . L'\'{e}quation de l'\'{e}quilibre est alors l'\'{e}quation
(1) (Serrin, 1986) o\`{u} }$\tau $ {\small s' interpr\`{e}te comme
le tenseur des containtes, }$\Omega ,\rho $ {\small et }$P$ {\small
d\'{e}signent respectivement le potentiel des forces de masse, la
densit\'{e} de masse et
la pression hydrodynamique. Si les forces de masse sont n\'{e}gligeables, l'%
\'{e}quilibre isotherme d'une bulle suppos\'{e}e sph\'{e}rique dans une
phase liquide de densit\'{e} }$\rho _{l\text{ }}${\small est repr\'{e}sent%
\'{e} par une solution de l'\'{e}quation (4). Pour chaque valeur }$\rho _{l%
\text{ }}${\small  dans une plage convenable on obtient un profil
croissant unique et une valeur de la densit\'{e} }$\rho _{v\text{ }}${\small %
au centre de la bulle. Le r\`{e}gle de Gibbs est donc bien v\'{e}rifi\'{e}%
e.\ Nous comparons l'\'{e}nergie libre d'un fluide homog\`{e}ne de densit%
\'{e} }$\rho _{o}\ ${\small contenu dans un domaine }${\small
D}${\small \
et l'\'{e}nergie libre de la m\^{e}me masse de fluide dans }${\small D}$%
{\small \ form\'{e}e par un liquide homog\`{e}ne contenant une petite bulle }%
${\small V}$. {\small  Cela nous conduit aux expressions (7) et (8) pour l'%
\'{e}nergie de nucl\'{e}ation }${\small W}${\small \ d'une bulle dans un
domaine infini, respectivement en th\'{e}orie de Laplace et en th\'{e}orie
du second gradient.\ Les conditions d'\'{e}quilibre liquide-vapeur en th\'{e}%
orie de Laplace permettent d'\'{e}crire }${\small W}${\small \ sous la forme
de l'\'{e}quation (11). La s\'{e}rie d'\'{e}galit\'{e}s (5), (6), (12),
(13), (14) montre que ce r\'{e}sultat se g\'{e}n\'{e}ralise sous la forme de
l'\'{e}quation (15) o\`{u} }$R_{m}^{2}$ {\small d\'{e}signe la valeur
moyenne de }$r^{2}${\small \ relativement \`{a} la mesure }$\rho ^{\prime 2}$%
{\small .\ En th\'{e}orie du second gradient, le tenseur des
contraintes est sph\'{e}rique \`{a} l'origine ainsi qu'\`{a}
l'infini mais, alors qu'\`{a}
l'infini la pression se confond avec la pression thermodynamique }$%
p_{l}=P(\rho _{l})${\small , cela est faux \`{a} l'origine o\`{u} }$%
p_{v}=P(\rho _{v})-C\rho _{v\ }\Delta \rho _{v\ }${\small . En identifiant
la diff\'{e}rence de pression entre l'origine et l'infini ainsi que l'\'{e}%
nergie de nucl\'{e}ation dans les deux th\'{e}ories, on obtient les d\'{e}%
finitions (17) et (18) pour le rayon }${\small R}${\small \ et la tension
superficielle }$\sigma ${\small \ de la bulle. Notre raisonnement est bas%
\'{e} sur l'\'{e}quation (1) obtenue par un mod\`{e}le de
m\'{e}canique des milieux continus. Il faut noter que van Kampen
(1964) a obtenu la m\^{e}me
\'{e}quation par des consid\'{e}rations de m\'{e}canique statistique. L'\'{e}%
quation (14) nous permet d'\'{e}valuer l'\'{e}nergie de
nucl\'{e}ation \`{a}
partir du profil de densit\'{e} donc de }$\rho _{l}${\small . La d\'{e}%
termination du rayon et de la tension de surface d'une bulle
microscopique par identification de l'\'{e}nergie de nucl\'{e}ation
et du saut de pression donn\'{e} par la th\'{e}orie de Laplace avec
les quantit\'{e}s correspondantes en th\'{e}orie du second gradient
n'a jamais \'{e}t\'{e} utilis\'{e}e dans la litt\'{e}rature: par
exemple Cahn (1959) et Evans (1979) ont utilis\'{e} une expression
de la tension superficielle valable pour une interface plane
uniquement et n'ont pas \'{e}tudi\'{e} la relation liant le rayon et
la tension superficielle. Les \'{e}quations (17) et (18) permettent
d'obtenir une telle relation sous forme implicite: le rayon et la
tension
superficielle d\'{e}pendent tous deux du profil de densit\'{e} qui lui-m\^{e}%
me d\'{e}pend de }$\rho _{l}${\small . Notre expression de la
tension de surface se r\'{e}duit, quand le rayon de la bulle
}${\small R}$ {\small  tend vers l'infini (\textit{i.e}. quand $\rho
_{l}$ tend }{\small   vers la valeur} $\rho _{l}^{P} ${\small
valable pour une interface plane), \`{a} l'expression usuelle (Cahn,
1959). L'\'{e}tude des variations de la tension
superficielle en fonction du rayon n\'{e}cessite le choix d'une loi d'\'{e}%
tat. Dell'Isola, Gouin et Rotoli (1996) ont   utilis\'e les potentiels chimiques du type de van der Waals propos%
\'{e}s par van Kampen (1964), Rocard (1967) ou Peng (1976), pour
effectuer des calculs num\'{e}riques: ils montrent que, pour des
bulles de rayon voisin du rayon critique, la
variation pr\'{e}vue de la tension superficielle correspond aux mesures exp%
\'{e}rimentales de Katz (1976) et Fisher (1980) alors que les
pr\'evisions de     Tolman (1947) et Kumar (1991) ne sont pas
r\'ealis\'ees.}

\end{document}